\documentstyle[12pt]{article}
\begin{document}
\large
\bibliographystyle{plain}
\begin{titlepage}
\large
\hfill\begin{tabular}{l}HEPHY-PUB 645/96\\ UWThPh-1996-30\\ July 1996
\end{tabular}\\[2.5cm]
\begin{center}
{\Large\bf RELATIVISTIC COULOMB PROBLEM:}\\[.5ex]
{\Large\bf ENERGY LEVELS AT THE CRITICAL}\\[.5ex]
{\Large\bf COUPLING CONSTANT ANALYTICALLY}\\
\vspace{1.8cm}
{\Large\bf Wolfgang LUCHA}\\[.5cm]
Institut f\"ur Hochenergiephysik,\\
\"Osterreichische Akademie der Wissenschaften,\\
Nikolsdorfergasse 18, A-1050 Wien, Austria\\[1.5cm]
{\Large\bf Franz F.~SCH\"OBERL}\\[.5cm]
Institut f\"ur Theoretische Physik,\\
Universit\"at Wien,\\
Boltzmanngasse 5, A-1090 Wien, Austria\\[2cm]
{\bf Abstract}
\end{center}
\normalsize
\noindent
The Hamiltonian of the spinless relativistic Coulomb problem combines the
standard Coulomb interaction potential with the square-root operator of
relativistic kinematics. This Hamiltonian is known to be bounded from below
up to some well-defined critical coupling constant. At this critical coupling
constant, however, the differences between all analytically obtainable upper
bounds on the corresponding energy eigenvalues and their numerically
determined (approximate) values take their maxima. In view of this, an
analytical derivation of (not so bad) upper bounds on the lowest-lying energy
levels at the critical coupling constant is presented.\\

\large
\end{titlepage}

The ``spinless relativistic Coulomb problem'' to be investigated here is
defined by a self-adjoint Hamiltonian $H$ composed of the square-root
operator of the relativistic expression for the free (so-called ``kinetic'')
energy
\begin{equation}
T\equiv\sqrt{{\bf p}^2 + m^2}
\label{eq:relkin1}
\end{equation}
of some particle of mass $m$ and momentum ${\bf p}$ as well as the
(spherically symmetric) Coulomb interaction potential
\begin{equation}
V_{\rm C}(r) = - \frac{\alpha}{r}\ ,\quad\alpha > 0\ ,
\label{eq:coulpot}
\end{equation}
depending merely on the radial coordinate $r \equiv |{\bf x}|$ and involving
some dimensionless coupling constant $\alpha$:
\begin{equation}
H = \sqrt{{\bf p}^2 + m^2} - \frac{\alpha}{r}\ .
\label{eq:ham-rcp}
\end{equation}

The serious investigation of the spectrum of the operator (\ref{eq:ham-rcp})
started with Ref.~\cite{herbst77}, where it was shown that, roughly speaking,
this operator is bounded from below precisely up to a critical value
$\alpha_{\rm c}$ of the coupling constant $\alpha$, namely, up to
$$
\alpha_{\rm c} = \frac{2}{\pi}\ .
$$
Unfortunately, until now this operator withstood all efforts to localize its
(discrete) spectrum analytically. (Some brief account of the history of these
attempts may be found, for instance, in Ref.~\cite{lucha94}.) In fact, from
the point of view of analytical statements about the eigenvalues of this
operator, at present one has to be content with sets of upper bounds on the
latter \cite{lucha94varbound,lucha96rcpaubel}.\footnote{\normalsize\ See also
Ref.~\cite{martin89}.} By comparing with numerical computations, however, one
realizes that, rather generally, the quality of all these upper bounds
decreases with increasing coupling constant $\alpha$. In view of the above,
in this note we are interested in the analytic calculation of upper bounds on
the eigenvalues of the Hamiltonian $H$ given in (\ref{eq:ham-rcp}) for large
values of the coupling constant $\alpha$, in particular, at the critical
coupling constant $\alpha_{\rm c}$.

The theoretical basis as well as the primary tool for the derivation of
rigorous upper bounds on the eigenvalues of some self-adjoint operator is,
beyond doubt, the so-called ``min--max principle'' \cite{reed-simon}. An
immediate consequence of this min--max principle is the Rayleigh--Ritz
technique: Let $H$ be some semibounded self-adjoint operator [which,
according to a comprehensive spectral analysis presented in
Ref.~\cite{herbst77}, obviously holds for our (semi-)relativistic
Hamiltonian, Eq.~(\ref{eq:ham-rcp})]. Let $E_k$, $k = 0,1,2,\dots$, denote
the eigenvalues of $H$, ordered according to $E_0\le E_1\le E_2\le\dots$. Let
$D_d$ be some $d$-dimensional subspace of the domain of $H$ and let $\widehat
E_k$, $k = 0,1,\dots,d-1$, denote all $d$ eigenvalues of this operator $H$
restricted to the space $D_d$, ordered according to $\widehat E_0\le\widehat
E_1\le\dots\le\widehat E_{d-1}$. Then the $k$th eigenvalue (counting
multiplicity) of $H$, $E_k$, satisfies the inequality
$$
E_k \le \widehat E_k\ ,\quad k = 0,1,\dots,d-1\ .
$$

Now, let us assume that this $d$-dimensional subspace $D_d$ is spanned by a
set of $d$ linearly independent basis vectors $|\psi_k\rangle$, $k =
0,1,\dots,d-1$. Then the set of eigenvalues $\widehat E$ may immediately be
determined as the $d$ roots of the characteristic equation
\begin{equation}
\det\left(\langle\psi_i|H|\psi_j\rangle -
\widehat E\,\langle\psi_i|\psi_j\rangle\right) = 0\ ,\quad
i,j = 0,1,\dots,d-1\ ,
\label{eq:chareq}
\end{equation}
as becomes clear from the expansion of any eigenvector of the operator $H$ in
terms of the basis vectors $|\psi_k\rangle$, $k = 0,1,\dots,d-1$, of the
subspace $D_d$.

The Coulombic Hamiltonian (3) involves only a single dimensional parameter,
namely, the particle mass $m$. Consequently, for a vanishing particle mass
all energy eigenvalues $E_k$ must vanish too, as is also seen by application
of the ``relativistic virial theorem'' proven in
Refs.~\cite{lucha89rvt,lucha90rvt}. Accordingly, in order to remain on the
safe side, we consider here only the special case of a nonvanishing particle
mass $m$:
$$
m > 0\ .
$$

Our particular choice for the basis vectors $|\psi_k\rangle$ to be adopted
here is defined by trial functions $\psi$ given in configuration-space
representation ($r \equiv |{\bf x}|$) by\footnote{\normalsize\ A proper
orthogonalization of this basis would yield basis functions which involve the
(generalized) Laguerre polynomials. In some respect, this orthogonal basis
appears to be more convenient for a numerical (which means, not entirely
analytical) treatment of the present problem \cite{lucha96num}.}
$$
\psi_k(r) = N_k\,r^{k+\beta-1}\exp(-m\,r)\ ,\quad\beta\ge 0\ ,
$$
with the normalization factor
$$
N_k = \sqrt{\frac{(2\,m)^{2\,k+2\,\beta+1}}{4\pi\,\Gamma(2\,k+2\,\beta+1)}}\ ,
$$
or in momentum-space representation ($p \equiv |{\bf p}|$) by
$$
\widetilde\psi_k(p) = \widetilde N_k\,\frac{\sin[(k+\beta+1)\arctan(p/m)]}
{p\,(p^2+m^2)^{(k+\beta+1)/2}}\ ,\quad\beta\ge 0\ ,
$$
with the normalization factor
$$
\widetilde N_k
= \sqrt{\frac{(2\,m)^{2\,k+2\,\beta+1}}{2\pi^2\,\Gamma(2\,k+2\,\beta+1)}}\,
\Gamma(k+\beta+1)\ ,
$$
both of which are, of course, related simply by Fourier transformation.
Whereas $k$ indicates a positive integer, $k = 0,1,2,\dots$, the parameter
$\beta$ is introduced to allow, for some given value of the coupling constant
$\alpha$, of a total cancellation of the divergent contributions to the
expectation values of kinetic energy $T$ and interaction potential $V_{\rm
C}(r)$: $\beta = \beta(\alpha)$. More precisely: in order to provide for that
cancellation, the parameter $\beta$ must be adjusted for the ground state
according to the relation \cite{cancel}
$$
\alpha = \beta\cot\left(\frac{\pi}{2}\,\beta\right)\ ,
$$
which implicitly determines $\beta$ as a function of the coupling constant
$\alpha$. In particular, this relation tells us that the critical coupling
constant, $\alpha_{\rm c}$, must be approached in the limit $\beta\to 0$.
From the behaviour of the matrix elements $\langle\psi_i|T|\psi_j\rangle$ of
the kinetic energy $T$ in Eq.~(\ref{eq:relkin1}) for large momenta $p$ and of
the matrix elements $\langle\psi_i|V_{\rm C}(r)|\psi_j\rangle$ of the Coulomb
interaction potential $V_{\rm C}(r)$ in Eq.~(\ref{eq:coulpot}) at small
distances $r$, respectively, it should become rather evident that, for our
choice of basis states $|\psi_k\rangle$, these singularities will arise
merely in those matrix elements which are taken with respect to
$|\psi_0\rangle$, that is, only in $\langle\psi_0|T|\psi_0\rangle$ and
$\langle\psi_0|V_{\rm C}(r)|\psi_0\rangle$.

The remainder of our way is straightforward to go. With the above basis
functions, we obtain for the matrix elements of the kinetic energy $T$ in
Eq.~(\ref{eq:relkin1})
$$
\begin{array}{l}\langle\psi_i|T|\psi_j\rangle\\[1ex]
= \displaystyle\frac{2^{i+j+2\,\beta+1}\,m}{\pi}\,
\displaystyle\frac{\Gamma(i+\beta+1)\,\Gamma(j+\beta+1)}
{\sqrt{\Gamma(2\,i+2\,\beta+1)\,\Gamma(2\,j+2\,\beta+1)}}\\[3ex]
\times\displaystyle\int\limits_0^\infty dy\,
\displaystyle\frac{\cos[(i-j)\arctan y] - \cos[(i+j+2\,\beta+2)\arctan y]}
{(1+y^2)^{(i+j+2\,\beta+1)/2}}\ ,\end{array}
$$
which may be evaluated with the help of the expansion
$$
\begin{array}{r}\cos(N\arctan y)
= \displaystyle\frac{1}{(1+y^2)^{N/2}}\,\displaystyle\sum_{n=0}^N
\left(\begin{array}{c}N\\n\end{array}\right)\cos\left(\frac{n\,\pi}{2}\right)
y^n\\[3ex]\mbox{for}\ N=0,1,2,\dots\ ,\end{array}
$$
for the matrix elements of the Coulomb interaction potential $V_{\rm C}(r)$ in
Eq.~(\ref{eq:coulpot})
$$
\langle\psi_i|V_{\rm C}(r)|\psi_j\rangle
= - \frac{2\,m\,\alpha\,\Gamma(i+j+2\,\beta)}
{\sqrt{\Gamma(2\,i+2\,\beta+1)\,\Gamma(2\,j+2\,\beta+1)}}\ ,
$$
and, finally, for the projections of the basis states $|\psi_k\rangle$ onto
each other
$$
\langle\psi_i|\psi_j\rangle = \frac{\Gamma(i+j+2\,\beta+1)}
{\sqrt{\Gamma(2\,i+2\,\beta+1)\,\Gamma(2\,j+2\,\beta+1)}}\ .
$$
Of course, some care has to be taken when extracting the singularity of the
matrix element $\langle\psi_0|T|\psi_0\rangle$, which may be done by
observing that
$$
\lim_{\beta\to 0}\,\int\limits_0^\infty dy\,
\frac{1 - \cos[(2+2\,\beta)\arctan y]}{(1+y^2)^{1/2+\beta}}
= 2\lim_{\beta\to 0}\,\int\limits_0^\infty dy\,
\frac{y^2}{(1+y^2)^{3/2+\beta}}\ .
$$
Introducing for notational simplicity a dimensionless and scaled energy
eigenvalue $\varepsilon$ by
$$
\widehat E = \frac{2}{\pi}\,m\,\varepsilon\ ,
$$
the characteristic equation for the Hamiltonian $H$ in (\ref{eq:ham-rcp})
thus becomes
$$
\det\left(\begin{array}{ccc}4\ln 2-2-\varepsilon&
\displaystyle\frac{\sqrt{2}}{3}-\displaystyle\frac{\varepsilon}{\sqrt{2}}&
\cdots\\[2ex]
\displaystyle\frac{\sqrt{2}}{3}-\displaystyle\frac{\varepsilon}{\sqrt{2}}&
\displaystyle\frac{17}{15}-\varepsilon&\cdots\\[2ex]
\vdots&\vdots&\ddots\end{array}\right) = 0\ .
$$
The roots of this characteristic equation are then given, for $d=1$, by
$$
\varepsilon = 2\,(2\ln 2 - 1)\ ,
$$
which entails the upper bound
$$
\frac{\widehat E_0}{m} = 0.4918\dots\ ,
$$
and, for $d=2$, by
$$
\varepsilon = \frac{1}{15}\left(60\ln 2 - 23 \pm
\sqrt{(60\ln 2)^2 -4800\ln 2 +1649}\right)\ ,
$$
which entails the upper bound
$$
\frac{\widehat E_0}{m} = 0.484288\dots\ ,
$$
for the ground-state eigenvalue $E_0$ of the Coulombic Hamiltonian
(\ref{eq:ham-rcp}). In principle, the $d$ (real) roots of the characteristic
equation (\ref{eq:chareq}) may be determined algebraically up to and
including the case $d=4$, entailing, of course, analytic expressions of
rather rapidly increasing complexity. For $d=4$, this yields for the
lowest-lying energy level the upper bound
\begin{equation}
\frac{\widehat E_0}{m} = 0.4842564\dots\ .
\label{eq:d=4upperbound}
\end{equation}

With respect to existing numerical determinations of energy levels, the
exploration of the Hamiltonian (\ref{eq:ham-rcp}) culminated so far with
Ref.~\cite{raynal94}, where---with the aid of the so-called local-energy
theorem for the lower bound and a standard variational procedure for the
upper bound---the admissible range of the energy eigenvalue of the ground
state has been narrowed down to an amazingly tiny gap. In particular, at the
critical coupling constant $\alpha_{\rm c}$ this range is numerically fixed
to be given by \cite{raynal94}
$$
0.4825\le\frac{E_0}{m}\le 0.4842910\quad\mbox{for}\ \alpha=\alpha_{\rm c}\ .
$$
Comparing these numerically determined bounds on the ground-state energy
level with our above upper bounds resulting algebraically from the
characteristic equation~(\ref{eq:chareq}), one immediately realizes that,
already for the case $d=2$, our analytical bound lies well within the
numerically obtained range. Consequently, the bound derived, in particular,
in the case $d=4$, Eq.~(\ref{eq:d=4upperbound}), represents, rather
obviously, a clear improvement of the best upper bounds known until now for
the ground-state energy eigenvalue of the spinless relativistic Coulomb
problem at the critical coupling constant. Needless to say, this simple
analytical investigation serves, in addition, to strengthen our confidence in
merely numerically computed upper bounds.

\end{document}